# A Pragmatic Machine Learning Approach to Quantify Tumor Infiltrating Lymphocytes in Whole Slide Images

Nikita Shvetsov[1], Morten Grønnesby[2], Edvard Pedersen[1], Kajsa Møllersen[3], Lill-Tove Rasmussen Busund[2,4], Ruth Schwienbacher[2,4], Lars Ailo Bongo[1], Thomas K. Kilvaer[5, 6,*]

[1] Department of Computer Science, UiT The Arctic University of Norway.
[2] Department of Medical Biology, UiT The Arctic University of Norway.
[3] Department of Community Medicine, UiT The Arctic University of Norway.
[4] Department of Clinical Pathology, University Hospital of North Norway.
[5] Department of Oncology, University Hospital of North Norway.
[6] Department of Clinical Medicine, UiT The Arctic University of Norway.
[*] Corresponding author: thomas.kilvaer@uit.no

## Abstract

**Motivation**

Increased levels of tumor infiltrating lymphocytes (TILs) in cancer tissue indicate favourable outcomes in many types of cancer. Manual quantification of immune cells is inaccurate and time consuming for pathologists. Our aim is to leverage a computational solution to automatically quantify TILs in whole slide images (WSIs) of standard diagnostic haematoxylin and eosin stained sections (H&E slides) from lung cancer patients. Our approach is to transfer an open source machine learning method for segmentation and classification of nuclei in H&E slides trained on public data to TIL quantification without manual labeling of our data.

**Results**

Our results show that additional augmentation improves model transferability when training on few samples/limited tissue types. Models trained with sufficient samples/tissue types do not benefit from our additional augmentation policy. Further, the resulting TIL quantification correlates to patient prognosis and compares favorably to the current state-of-the-art method for immune cell detection in non-small lung cancer (current standard CD8 cells in DAB stained TMAs HR 0.34 95% CI 0.17-0.68 *vs* TILs in HE WSIs: HoVer-Net PanNuke Aug Model HR 0.30 95% CI 0.15-0.60, HoVer-Net MoNuSAC Aug model HR 0.27 95% CI 0.14-0.53). Moreover, we implemented a cloud based system to train, deploy and visually inspect machine learning based annotation for H&E slides. Our pragmatic approach bridges the gap between machine learning research, translational clinical research and clinical implementation. However, validation in prospective studies is needed to assert that the method works in a clinical setting.

**Availability and implementation**
Web server repository: https://github.com/uit-hdl/histology

Machine learning pipeline: https://github.com/uit-hdl/hovernet-pipeline and convenience tools for easy setup: https://github.com/uit-hdl/hover_build
Publicly available datasets that we used to train our models:
https://warwick.ac.uk/fac/cross_fac/tia/data/hovernet (ConSep),
https://warwick.ac.uk/fac/cross_fac/tia/data/pannuke (PanNuke) and
https://monusac-2020.grand-challenge.org/Data/ (MoNuSAC)
UiT-TILs dataset and trained models: https://doi.org/10.18710/4YN9SZ
Open access web server: https://hdl-histology-ne.azurewebsites.net/

**Contact**
thomas.k.kilvar@uit.no

# Introduction

Increasing availability of digital pathology opens new possibilities. Digital whole slide images (WSIs) stored on servers are easily retrieved for review by pathologists. Viewing serial WSIs from the same patient side-by-side enables the pathologist to assess morphology and protein expression at the same time, as opposed to swapping the slides back-and-forth when viewing in a microscope. As a bonus, the field of view is vastly increased compared to the 1000μm circle offered at 400x in a microscope. Further, residents in pathology may demark areas of interest and pass these to consultant pathologists for feedback and learning.

Although digital pathology will optimize current workflows, computational pathology is by many believed to be one of its most significant advantages. A standard three channel WSI is approximately 100 000 x 100 000 pixels. This represents an enormous amount of information that may be used to identify, and subsequently quantify, macro- (patches of cartilage and bone, islets of cancer or vessels), intermediate- (different types of cells) and micro-structures (different cellular components). Whilst some of this information is utilized in classical light microscopy, important information is likely discarded. Moreover, additional image features, abstract to the human mind, may be extracted and incorporated as biomarkers into existing diagnostic pipelines expanding the existing prognostic and predictive toolkit. An additional benefit may be a reduction of intra- and interobserver variations (Jackson *et al.*, 2017) - a known diagnostic challenge that is especially pronounced if the problem is complex and/or open for personal opinion.

A hematoxylin & eosin (H&E) stained tissue slide is the principal method used to evaluate tissue morphology, while immunohistochemistry (IHC), with chromogens or fluorophores bound to an antibody, is used to visualize specific protein expressions. Classically, focus has been on the morphology and protein expression of cancer cells. However, numerous studies highlight the interactions between cancer cells and their surroundings. This interplay, popularly termed the tumor microenvironment, impacts patients' prognoses and is likely important when deciding treatment strategies. Immune cells, and especially tumor infiltrating lymphocytes (TILs), are among the most important cells in the tumor microenvironment and will in many cases directly influence cancer development. As a result, different variations of TILs have been suggested as prognostic and/or predictive biomarkers in various types of cancer including colorectal cancer, breast cancer, melanoma and non-small cell lung cancer (NSCLC) (Hendry *et al.*, 2017; Salgado *et al.*, 2015; Pagès *et al.*, 2018; Donnem *et al.*, 2015). Interestingly, strategies for TIL identification vary between cancer types. In colorectal cancer (CRC), IHC is used to visualize CD3+ and CD8+ TILs (Pagès *et al.*, 2018), while conventional H&E is used for TIL identification in breast cancer and melanoma (Salgado *et al.*, 2015; Hendry *et al.*, 2017). Moreover, TIL quantification ranges from fully discrete grouping, *via* semi-quantitative to absolute count based methods. For other cancers, such as NSCLC, the preferred method of TIL identification and quantification is yet to be determined (Hendry *et al.*, 2017; Donnem *et al.*, 2015, 2016). While IHC based methods provide information about cell type and function, they add complexity, time spent and cost. Hence, using conventional H&E for TIL identification is tempting - especially if identification can be conducted without time-consuming manual counting.

Computational pathology presents an opportunity for automatic TIL identification in H&E WSIs. The methods range from simple rule based systems, which rely on easily understandable hand crafted features like shape, size and color, to deep learning (DL) approaches, which calculate and combine image features non-linearly. Given proper implementation, DL will usually outperform the sensitivity/specificity of simpler approaches, at the cost of increased complexity, calculation time and loss of interpretability. Unfortunately, DL requires thousands of annotated images to reach full potential and procurement of annotated training datasets may be challenging due to the required pathologist effort. Fortunately, several recent projects provide data-sets and introduce highly accurate DL models for cell segmentation and classification in H&E images (Graham *et al.*, 2019; Vu *et al.*, 2019; Sirinukunwattana *et al.*, 2016; Gamper *et al.*, 2019; Kumar *et al.*, 2017; Verma *et al.*, 2020). However, to our knowledge, none of these approaches have been adapted to TIL quantification in a clinical setting.

Herein, we present our approach and cloud-based system for quick evaluation and deployment of machine learning models in a clinical pathology setting using our real world dataset from patients with NSCLC. To investigate the clinical usefulness of DL for TIL quantification we have:

1. Implemented the HoVer-Net (Graham et al., 2019) algorithm and compared our results to the original paper and the work of (Gamper *et al.*, 2019).
2. Tested how HoVer-Net performs on unseen data, which will be the case in a clinical setting.
3. Evaluated an additional augmentation approach with the aim of improving performance on unseen data (Balkenhol *et al.*, 2018).
4. Compared the prognostic performance of TILs identified by HoVer-Net to the standard state-of-the-art approach (CD8 staining).
5. Assessed the immune cell annotations produced by six HoVer-Net models trained on three datasets for our unlabeled data set comprising H&E stained images of NSCLC.
6. Created a system for deploying machine learning models in a cloud to support the implementation of computational pathology, and to quickly evaluate promising methods in the clinical setting.

# Material and Methods

We present our approach and cloud-based system for transferring open source machine learning models trained on public datasets to solve new problems in H&E WSI annotation. As a proof-of-concept, we customize the HoVer-Net algorithm (Graham *et al.*, 2019) to detect and quantify TILs in tissue samples from patients diagnosed with NSCLC. However, the developed pipeline is easily adapted to any object detection task in WSIs. To validate the customized models performances on our data without annotating hundreds of thousands of cells, we use visual inspection of inference results and confirm their clinical utility in 1189 1000x1000µm patches at a resolution of 0.2428µm/px from 87 NSCLC patients.

## Datasets

### Public datasets to train and test a machine learning model

In the past decade, several excellent annotated datasets for cell segmentation and classification were made publicly available as summarized in Table 1. The CoNSeP, PanNuke and MoNuSAC datasets are at the time of submission the only publicly available datasets comprising both the segmentation masks and class labels necessary to train DL models for simultaneous cell segmentation and classification. Dataset characteristics are summarized in table 1 and a simplified version of how patches are generated and processed during the training procedure is illustrated in figure 1.

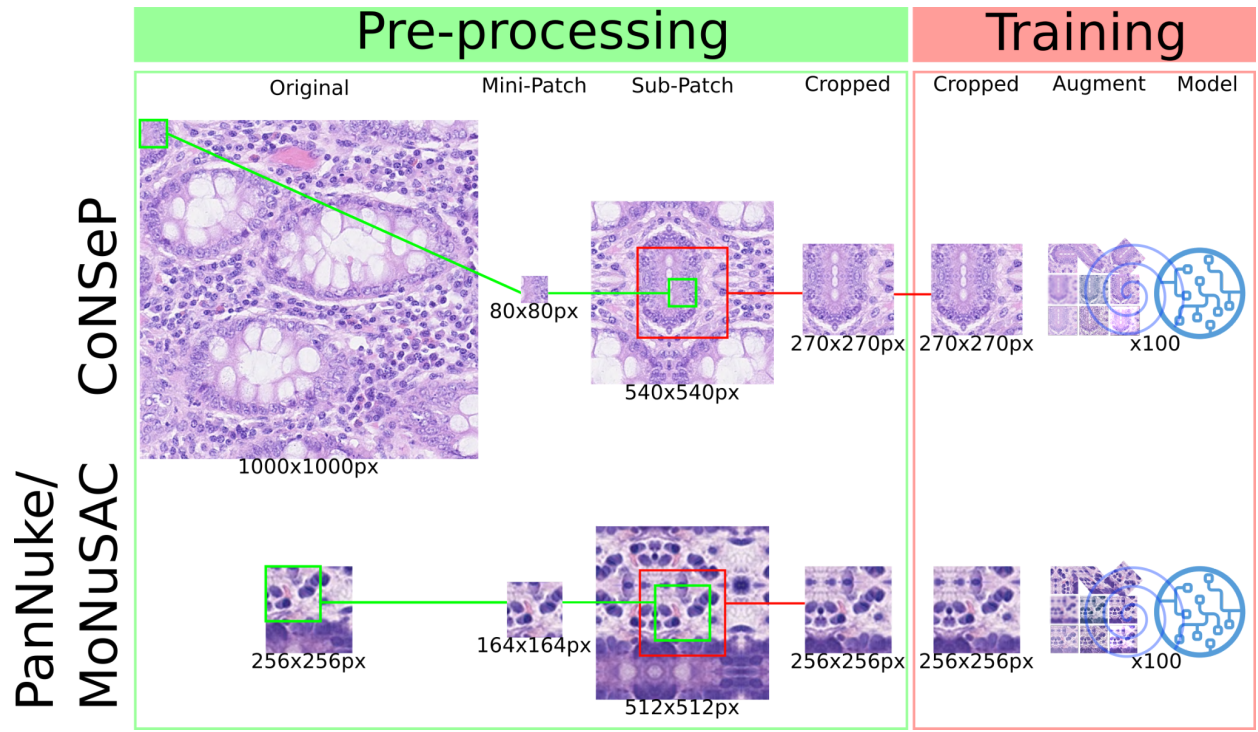

**Figure 1:** For each dataset, the original patches are divided into non-overlapping (CoNSeP) and overlapping (PanNuke/MoNuSAC) mini-patches (green squares). A sub-patch is generated by adding padding using image information from adjacent tissue in the original image patch and/or by mirroring in edge cases. The central part of the sub-patch is cropped and used in the training procedure (red square). The output of the network (segmented/classified cells) is equal to the mini-patch (green square). Each model is subsequently trained for 100 epochs.

**Table 1**: Characteristics of datasets providing annotations and/or classifications publically available for training and validation of models for instant cell segmentation and classification.

| Dataset | CoNSeP | PanNuke | MoNuSAC | CRCHisto* | TNBC | MoNuSeg | CPM-15 | CPM-17 |
|---|---|---|---|---|---|---|---|---|
| Author | Graham | Gamper | Verma | Sirinukunwattana | Naylor | Kumar | Vu | Vu |
| Year | 2019 | 2020 | 2020 | 2019 | 2017 | 2017 | 2019 | 2019 |
| Origin | UHCW | UHCW/TCGA | TCGA | UHCW | Curie Institute | TCGA | TCGA | TCGA |
| Tissue types | CRC | Various (19) | Various (4) | CRC | TNBC | Various (8) | Various (2) | Various (4) |
| Unique patients | | | | | | | | |
| Number of patches | 41 | 7901 | 294 | 100 | 50 | 30 | 15 | 64 |
| Training | 27§ | 2722 | 209 | NA | NA | 16# | NA | 32 |
| Validation | § | 2656 | NA | NA | NA | # | NA | NA |
| Testing | 14 | 2523 | 85 | NA | NA | 14# | NA | 32 |
| Patch size | 1000x1000 | 256x256 | Various | 500x500 | 512x512 | 1024x1024 | 400x400 up to 600x1000 | 500x500 up to 600x600 |
| Scanner(s) | Omnyx VL120 | Various | Various | Omnyx VL120 | Philips Ultra Fast Scanner 1.6RA | Various | Various | Various |
| Magnification | 40x | 40x | 40x | 20x | 40x | 40x | 40x and 20x | 40x and 20x |
| Resolution | 0.275µm/px | Various | NA | 0.55µm/px | 0.245µm/px | NA | NA | NA |
| Annotation | NC | NC | NC | CoN | NC | NC | NC | NC |
| Cells | 24 319 | 205 343 | 46 000 | 29 756 | 4 022 | 21 623 | 2 905 | 7 570 |
| Labeled cells | 24 319 | 205 343 | 46 000 | 22 444 | NA | NA | NA | NA |
| Cell types | 7 | 5 | 4(5) | 4 | NA | NA | NA | NA |

Abbreviations: CoNSeP, ; CRCHisto, colorectal histology; PanNuke, ; MoNuSAC, ;TNBC, triple negative breast cancer; MoNuSeg, ; CPM, ; NC, nucleus contour; UHCW, University Hospital; TCGA, The Cancer Genome Atlas; CiN, centre of nucleus; NG, not given.
*CRCHisto only provides centre of nucleus annotations
§(Graham *et al.*, 2019) suggests to split the 27 training images into a training and validation set. However, the paper does not provide information on the split conducted in their published paper.
#(Kumar *et al.*, 2017) suggests to split into a train/validation set of 16 and 14 images. They also provide the split conducted in their published paper.

## UiT-TILs dataset used to clinically validate TIL classifications

To validate the clinical relevance of TIL detection, we compiled the *UiT-TILs* dataset comprising 1189 image patches from 87 NSCLC patients with matched clinical data. The *UiT-TILs* dataset is a subset of the cohort presented by Rakaee et al. in 2018 (Rakaee *et al.*, 2018). The patches were generated in QuPath (Bankhead *et al.*, 2017) using thresholding to detect the tissue area, followed by generation of 4019x4019px (1000x1000µm) patches and manual selection of up to 15 consecutive patches from the tumor border. Due to differences in tumor size, tissue quality and the absolute length of the tumor border present in each WSI, the target number of 15 patches from each patient could not be met in all instances (median number = 15, range 3 - 16). The patient's clinical data are summarized in Table S1.

## Augmentation

Our augmentation strategy includes the use of a heuristic data augmentation scheme thought to boost classification performance (Ratner *et al.*, 2017). First, we augment the two training datasets using the augmentation policies provided in the HoVer-Net source code (Graham *et al.*, 2019). These include affine transformations (flip, rotation, scaling), Gaussian noise, Gaussian blur, median blur and color augmentations (brightness, saturation and contrast). Second, we introduce a linear augmentation policy proposed by Balkenhol et al. (Balkenhol *et al.*, 2018) for better generalizability and overall improvement of trained models. The image is, 1) converted from the RGB to the HED color space using the scikit-image implementation of *color deconvolution* according to the Ruifrok and Johnston algorithm (Ruifrok and Johnston, 2001), 2) each channel is transformed using a linear function which stochastically picks coefficients from a predefined range, and 3) transformed channels are combined and converted back to the RGB space.

# Replicated training algorithm

We repurpose the training pipeline from the HoVer-Net source code. *In brief,* we preprocess the data and extract sub-patches from the training dataset. For the CoNSeP, MoNuSAC and PanNuke datasets the input patch sizes are 270x270px, 256x256px and 256x256px, respectively. For each input patch, RGB channels are normalized between 0 and 1. The datasets are augmented via predefined policies and handled by the generator during training. Figure 1 is a simplistic depiction of pre-processing and training. The training procedure is performed in two phases - training decoders, while freezing encoders and fine-tuning of the full model. The training is initialized on weights pre-trained on ResNet50 and optimized using Adam optimization with decreasing learning rate after 25 epochs for each phase. The models are trained for a total of 100 epochs. The ConSeP dataset used in Graham et al is split into train and test subsets with 27 and 14 image patches respectively. Since no validation set is provided for the CoNSeP dataset, we use the training and test sets to train and validate the CoNSeP models and the test set is used again to evaluate model performance. The MoNuSAC dataset provides train and test sets, when training our models we split the train set 80/20 into train and validation sets respectively. The PanNuke dataset provides separate sets for training, validation and testing models. For additional details about the training parameters we refer to the original paper and their open source code (Graham and Vu; Graham *et al.*, 2019; Verma *et al.*, 2020).

# A system for training, tuning and deploying machine learning models

To easily tune and evaluate machine learning models we need an environment that has the compute resource to keep the training time short, provide easy deployment, keep track of experiment configuration and ensure that our experiments are reproducible. In addition, we want to make the code portable, since we want to deploy the containers on a commercial cloud with GPU support to get access to the needed compute resources for fast training, or in a secure platform to train on sensitive data that cannot be moved to a commercial cloud.

To solve the above challenges, we containerize the code so that we can easily deploy it on a cloud platform or our in-house infrastructure. We use Docker (Merkel, 2014) containers, and a Conda (Anaconda Software Distribution, 2020) environment for dependency handling. We use Kubernetes (Cloud Native Computing Foundation) for container management. To ensure experiment reproducibility and to keep track of tuning parameters we use Pachyderm (Joey Zwicker, Daniel Whitenack, Joe Doliner) since it enables data and model provenance and parallelized batch processing.

We deploy our customized HoVer-Net model using our approach summarized in Figure 2. To make it easier to test different training data, augmentation, and parameters we use configuration profiles implemented as YAML files, generated in the process and stored in the code repository. We split data wrangling and training from inference to make it easier to deploy these on a new platform. To reduce execution time, we use Tensorpack with Tensorflow 1.12 to infer and post-process patches in parallel.

We use a docker container on a local server with a RTX Titan GPU for development and debugging. We train and tune the models on the Azure cloud, using AKS (Azure Kubernetes

Service) with GPU nodes on top of the Pachyderm framework. As a node pool we use 1 Standard NC6 (6 cores, 56 GB RAM) node with 1xK80 GPU accelerator.

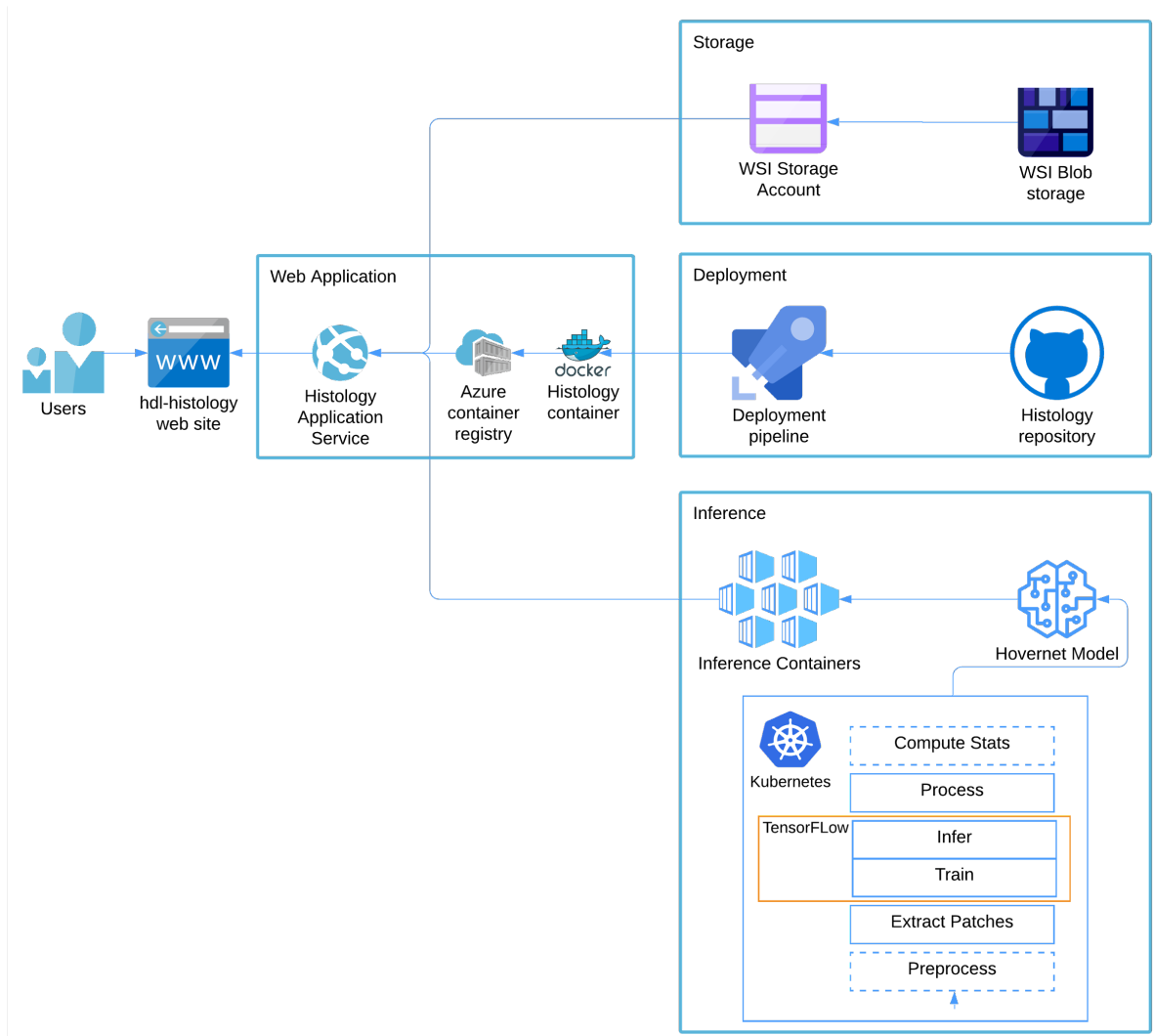

**Figure 2:** Overview of cloud deployment of the web application. WSIs are uploaded into Azure Blob storage, and mapped in as a file system to the web application. The web application is built by an Azure Pipeline from a docker file contained in the Histology git repository. We use the Kubernetes platform in Azure to train and validate our implementation of HoVer-Net. The exported models are implemented as inference containers in the Histology application service and served using tensorflow serving.

## Interactive annotation of WSIs

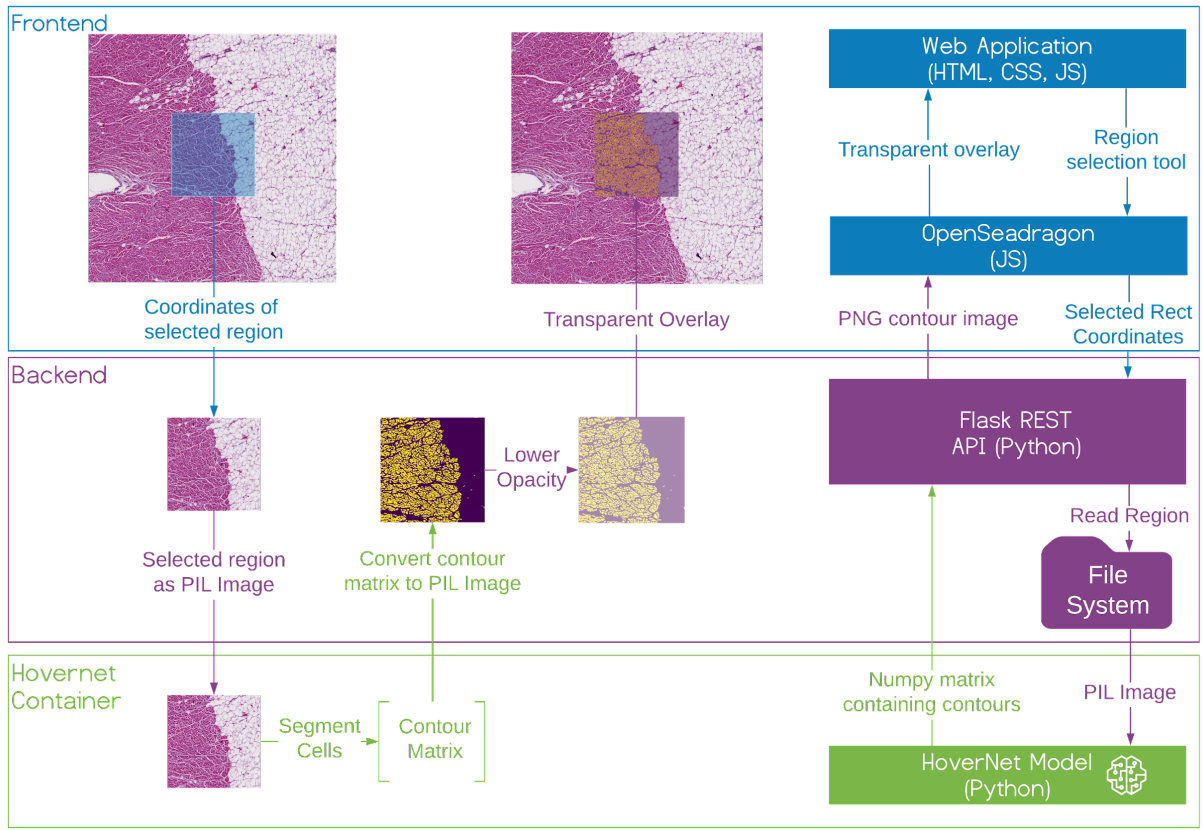

**Figure 3**: Overview of the slide viewer frontend, backend and inference container. Selected regions of the frontend are cropped out at the backend and transferred to the inference container. The inference container returns a contour matrix which is converted into a contour image and overlaid on the frontend WSI.

  To demonstrate and visually evaluate the TIL classifications and quantification, we developed the hdl-histology system (Figure 3). It is an interactive web application for exploring and annotating gigabyte-sized WSIs. Our web server implements the URL endpoints required to request and transfer WSI tiles with interactive response times. It uses the Flask framework to implement a RESTful API. We use Gunicorn (Green Unicorn) (Benoit *et al.*) to handle concurrent requests. We use OpenSeadragon (OSD) to implement a tiled map service (The OpenSeadragon Authors.). To achieve interactive performance for high resolution WSIs, we export the slides from their native format (.vms, .mrxs, .tiff etc) using *vips dzsave,* and generate Microsoft's DeepZoom Image (.dzi) tiles as a pyramidal folder hierarchy of .jpeg's on disk. We provide classification as a service for interactive patch annotation. We use Tensorflow Serving as a containerized server for trained models. This service exports a REST API. The input is the patch to classify and the name of the model to use. The service returns a class probability map that can be used to create a visual mask for the cells in the patch. The REST API is used by the webserver that overlays the returned pixel matrix of annotated regions.

  The web server runs in a docker container on the Azure App Service. The App Service pulls the Docker Image from our Azure Container Registry. The images are stored in a Blob Container which is path mapped to our App Service. To deploy updates to the system, we use continuous integration with Azure Pipelines. The deployment pipeline triggers on push events to an azure branch in the git repository using Github Webhooks (https://docs.github.com/en/developers/webhooks-and-events/about-webhooks). The build pipeline pulls the latest change from the branch and builds a docker container from the repos Dockerfile and pushes the container image to our Azure Container Registry (Figure 2).

## Validation of TIL quantification on unlabeled data

To pragmatically evaluate the quantification of TILs without the large effort required for manual labeling we apply three methods.

*First*, to verify that our modifications to HoVer-Net do not change the results, we reproduce the results described in the original HoVer-Net and PanNuke papers using the CoNSeP and PanNuke datasets (S2 table, (Gamper *et al.*, 2020; Graham *et al.*, 2019). Further, to investigate whether additional augmentation leads to more robust and generalizable models, we use the CoNSeP, MoNuSAC and PanNuke datasets to train separate models, with and without additional augmentations.

For comparison to the original HoVer-Net paper, we evaluate the segmentation and classification tasks using the metrics provided in the HoVer-Net source code (Graham *et al.*, 2019). The full description of metrics used in the replication study is available in the supplementary materials.

For all other comparisons we use conventional and well established metrics to compare and benchmark our models. Since the models encompass both segmentation and classification, we compare segmentation and classification separately and combined. The definitions of the metrics we used to evaluate segmentation and classification separately are available in the supplementary materials. Integrated segmentation/classification metrics include:

$$Accuracy_{dc}^t = \frac{TP_{dc}^t + TN_{dc}^t}{TP_{dc}^t + TN_{dc}^t + FP_{dc}^t + FN_{dc}^t}, Precision_{dc}^t = \frac{TP_{dc}^t}{TP_{dc}^t + FP_{dc}^t}, Recall_{dc}^t = \frac{TP_{dc}^t}{TP_{dc}^t + FN_{dc}^t},$$

$$F1_c^t = \frac{TP_{dc}^t}{TP_{dc}^t + 0.5(FP_{dc}^t + FN_{dc}^t)}, \text{ where}$$

$TP_{dc}^t$ - detected cells with GT label t, classified as t,

$TN_{dc}^t$ - detected or falsely detected cells with GT label other than t, classified as other than t,

$FP_{dc}^t$ - detected or falsely detected cells with with GT label other than t, classified as t,

$FN_{dc}^t$ - detected cells with GT label t, classified as other than t and all cells with GT label t not detected, for each class t.

As $t$, we use inflammatory and cancer cells. To compare our augmentation approach with the original approach we applied the trained models to their respective test datasets and used cross-inference with test data from the other dataset. For evaluation of cross-inference results we had to count both normal epithelial and neoplastic cells as cancer cells, due to the fact that the CoNSeP and MoNuSAC datasets do not differentiate between normal and neoplastic epithelial cells.

*Second*, to demonstrate that our methods may be adapted for use in a clinical setting, we apply our models on all patches in the UiT-TILs dataset and calculate the median number of cells classified as inflammatory/TILs for each patient. The median number of identified immune cells wilø then be correlated to the numbers of immune cells identified using digital pathology and specific immunohistochemistry on tissue micro-arrays, and semi-quantitatively scored TILs previously evaluated in the same patient cohort (Kilvaer *et al.*, 2020; Rakaee *et al.*, 2018). To assess the impact of TILs identified using our deep learning approach, patients are stratified into high and low TIL groups and their outcome is compared using the Kaplan-Meier method and the log-rank test with disease-specific survival as endpoint.

*Third*, to demonstrate that our methods are meaningful in a clinical pathology setting, two experienced pathologists (Busund, Schwienbacher) visually inspected the segmentation and classification results for TILs and cancer cells in 20 randomly sampled patches from the provided dataset. During the inspection they estimated the precision and recall for TILs and cancer cells using 10% increments for each measurement. We calculate F1 scores based on the estimated precision and recall for each patch.

# Results

## Model performance

We trained a total of three pairs of HoVer-Net models with and without additional augmentations using training data from: 1) CoNSeP, 2) MoNuSAC and 3) PanNuke. Each model's performance was evaluated on its respective test set and on the test set of the other models. Additional information on training and inference times are available in the supplementary material.

*First*, we verified that our replication of the original HoVer-Net models, trained on the CoNSeP and PanNuke datasets, provided comparable performance to the results presented by Graham et al. (Graham et al., 2019) and Gamper et al. (Gamper *et al.*, 2020). The results are presented in S2 Table and show that our implementations perform within the expected range compared to the original studies' results. We believe the small differences are due to missing details in the original studies (for ConSeP the partitioning and preprocessing is not clearly described; for PanNuke the source code and model hyperparameters are not available).

*Second*, to get an estimation on real-world model performance, we evaluated our original HoVer-net models trained on CoNSeP, MoNuSAC and PanNuke data on the corresponding test sets for each dataset. As expected, all models suffered a drop in performance when used for inference on a test set generated from another data source. This effect was most prominent for the model trained on CoNSeP data, whose performance dropped significantly when tested on MoNuSAC and PanNuke data (Tables 2 and S3, columns BI and CI *vs* AII). The performance drop was especially prominent in classification and was consistent for both cancer and inflammatory cells. A similar trend was observed for the model trained on MoNuSAC data (Tables 2 and S3, columns CV *vs* AV and BV). The PanNuke model tested on ConSeP and MoNuSAC data (Tables 2 and S3, column BIII *vs* AIII and CIII) exhibited less variable results on different data.

*Third*, we evaluate the effect of our augmentation policies on cross-inference performance since better augmentation often improves model transferability. However, our results do not indicate a significant performance improvement neither for the CoNSeP model (Table 2, columns B I *vs* B II and CI *vs* CII), the MoNuSAC model (Table II, columns AV *vs* AVI and BV *vs* BVI) nor the PanNuke model (Table 2, columns AIII *vs* AIV and CIII *vs* CIV).

Finally, we compare the augmented models (Table 2). The CoNSeP and MoNuSAC models perform best on their own data and experience performance drops when used on unseen data. Interestingly, the PanNuke model performs better on MoNuSAC data and compares favorably with models trained on the CoNSeP and MoNuSAC datasets when tested on their data. Consequently, we believe the PanNuke test data results are most representative for the expected performance on the lung tumor tissue in the UiT-TILs data, since PanNuke includes patches with this tissue type. We therefore expect PanNuke to perform better on UiT-TILs.

**Table 2**: A summary of the performance of four deep learning models trained using the CoNSeP (A I-II and B I-II) and the PanNuke (A III-IV and B III-IV) datasets using the original training pipeline as published by Graham et al. (Graham *et al.*, 2019) without (A I and III and B I and III) and with (A II and IV and B II and IV) enhanced augmentation (Graham *et al.*, 2019). The best results for each parameter 1) within each dataset are in bold and 2) for models trained on another dataset are in italics. Separate comparisons of the segmentation and classification steps are provided in S3 Table.

| Test data | CoNSeP | | | | | | PanNuke | | | | | | MoNuSAC | | | | | |
|---|---|---|---|---|---|---|---|---|---|---|---|---|---|---|---|---|---|---|
| Model | CoNSeP | | PanNuke | | MoNuSAC | | CoNSeP | | PanNuke | | MoNuSAC | | CoNSeP | | PanNuke | | MoNuSAC | |
| Augmentation | HoVer | Aug | HoVer | Aug | HoVer | Aug | HoVer | Aug | HoVer | Aug | HoVer | Aug | HoVer | Aug | HoVer | Aug | HoVer | Aug |
| Numbering | AI | AII | AIII | AIV | AV | AVI | BI | BII | BIII | BIV | BV | BVI | CI | CII | CIII | CIV | CV | CVI |
| Integrated segmentation and classification | | | | | | | | | | | | | | | | | | |
| $Accuracy_{dc}^{inflammatory}$ | 0.78 | **0.80** | 0.71 | 0.71 | *0.72* | 0.55 | 0.73 | **0.76** | 0.71 | 0.72 | 0.47 | 0.46 | 0.59 | 0.70 | 0.74 | **0.75** | 0.73 | 0.73 |
| $Precision_{dc}^{inflammatory}$ | 0.81 | 0.77 | 0.60 | 0.70 | ***0.84*** | 0.79 | *0.59* | 0.52 | 0.61 | **0.68** | 0.48 | 0.45 | 0.86 | 0.79 | 0.78 | ***0.92*** | 0.77 | 0.77 |
| $Recall_{dc}^{inflammatory}$ | 0.58 | 0.66 | 0.79 | 0.74 | ***0.83*** | 0.60 | 0.28 | 0.36 | **0.64** | 0.61 | 0.53 | *0.57* | 0.26 | 0.49 | 0.78 | *0.80* | 0.86 | **0.87** |
| $F1_{dc}^{inflammatory}$ | 0.68 | 0.71 | 0.68 | 0.72 | ***0.83*** | 0.68 | 0.38 | 0.42 | 0.63 | **0.64** | *0.50* | *0.50* | 0.40 | 0.60 | 0.78 | ***0.86*** | 0.81 | 0.82 |
| $Accuracy_{dc}^{cancer}$ | 0.57 | **0.59** | *0.54* | 0.51 | 0.38 | 0.38 | 0.42 | *0.45* | 0.57 | **0.58** | 0.45 | 0.45 | 0.53 | *0.63* | *0.63* | 0.62 | 0.68 | **0.70** |
| $Precision_{dc}^{cancer}$ | 0.59 | 0.61 | 0.61 | 0.62 | 0.68 | **0.72** | 0.53 | 0.50 | 0.64 | 0.63 | 0.70 | ***0.72*** | *0.70* | 0.62 | 0.58 | 0.57 | 0.71 | **0.73** |
| $Recall_{dc}^{cancer}$ | **0.74** | 0.73 | *0.68* | 0.66 | 0.45 | 0.44 | 0.26 | 0.34 | 0.68 | **0.73** | *0.51* | 0.50 | 0.10 | 0.42 | 0.92 | ***0.93*** | 0.84 | 0.85 |
| $F1_{dc}^{cancer}$ | **0.66** | 0.66 | *0.64* | *0.64* | 0.54 | 0.55 | 0.35 | 0.40 | 0.66 | **0.68** | *0.59* | *0.59* | 0.18 | 0.50 | *0.71* | *0.71* | 0.77 | **0.79** |

## Correlation of TIL quantification with cancer survival rate

We compared the number of TILs identified with the PanNuke aug model with the number of TILs identified using our simple rule based approach (*Helm*, described in Supplementary materials), HoVer-Net models trained on the CoNSeP, MoNuSAC and PanNuke datasets with and without additional augmentations and different subsets of T-lymphocytes in TMAs previously investigated in the same patients (Kilvaer *et al.*, 2020) (S1 Figure). TILs identified using the rule based approach (R 0.44) and with the CoNSeP (Original R 0.54; Aug R 0.92), MoNuSAC (Original R 0.89; Aug R 0.90) and PanNuke models (Original R 0.94) were moderately to strongly correlated with TILs identified with the PanNuke aug model. Moderate correlations were also observed with the number of TILs identified by the pan T-lymphocyte marker CD3 and the cytotoxic T-cell marker CD8 (R-values of 0.54 and 0.39, respectively) in TMAs from the same patients.

We evaluated the clinical impact of TILs identified using our approach in the UiT-TILs cohort. The results are summarized in Figure 4 and in S4 Table. In brief, we show that TILs, identified by either the original and augmented MoNuSAC and PanNuke models or the augmented CoNSeP model, can be used to identify TILs as prognostic factors in NSCLC (S4 Table and Figure 4 panel E-G). Specifically, patients with an above median number of TILs present with superior survival and compared favorably with state-of-the-art CD8 lymphocyte detection in the same patients (S4 Table and Figure 4 panel C). This shows that a deep learning model could be used instead of CD8 staining to quantify TILs.

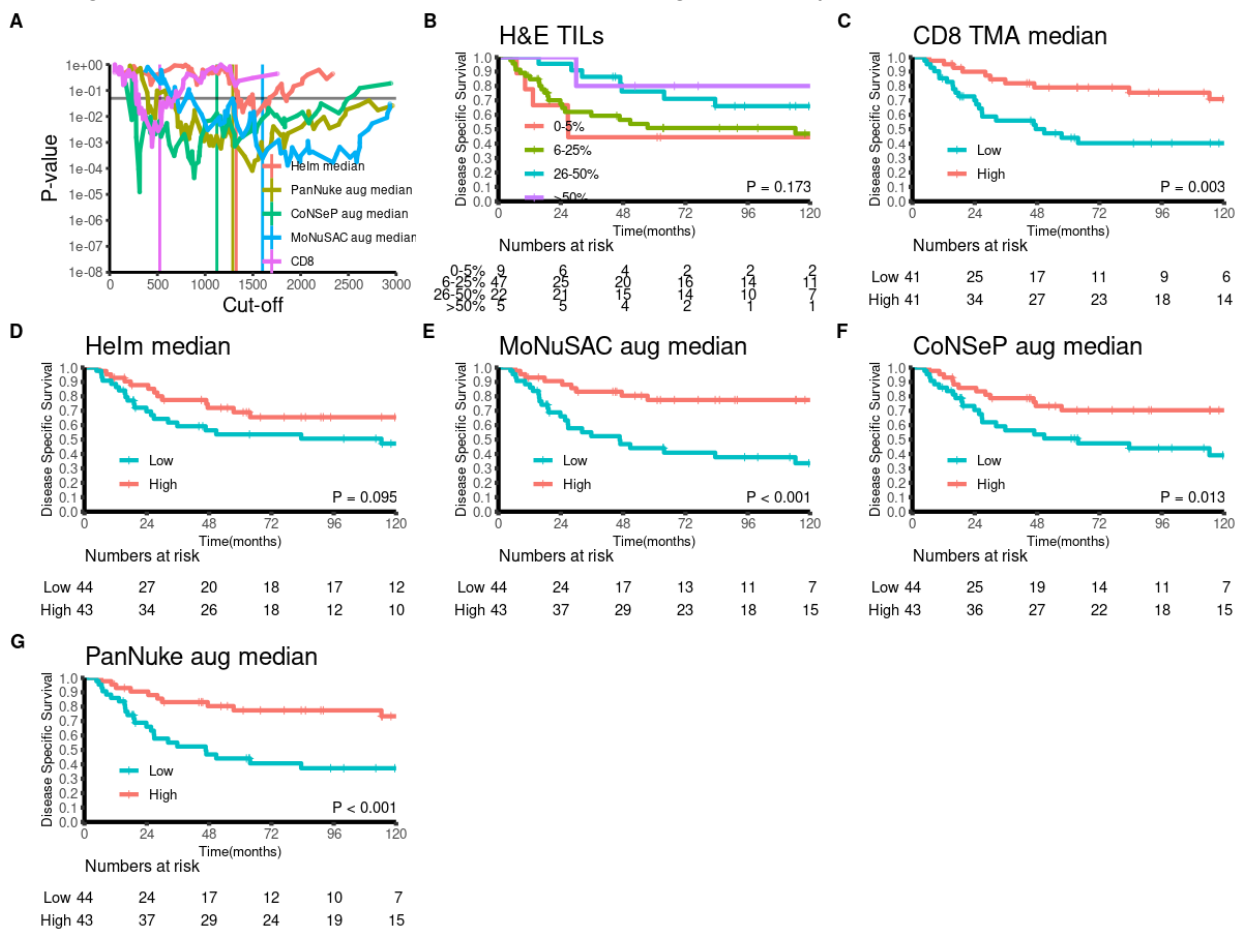

**Figure 4:** Panel A) All possible dichotomized cut-offs for TILs identified by CD8 IHC in TMAs or using the Helm, PanNuke or CoNSeP models plotted against P-values indicating significance of DSS for all included patients (N = 87). Panels B-G) Disease-specific survival curves for high and low TILs scores for B) the semi-quantitative model proposed in (Rakaee *et al.*, 2018); C) the CD8 model proposed in (Kilvaer *et al.*, 2020); D-F) high and low number of TILs identified using the baseline rule based Helm algorithm (described in Supplementary materials) or DL models trained on the CoNSeP, MoNuSAC, PanNuke datasets, respectively.

## Manual validation of classification results

Two pathologists, LTRB and RS, manually validated the performance of three original and three augmented HoVer-Net models in a real world dataset comprising 20 patches. Results for all patches are provided in Tables S5 and S6. For cancer cells, recall was > 90% for all models (PanNuke original: median 97%, range 96 - 97%, PanNuke aug: median 97%, range 97 - 97%, CoNSeP original: median 97%, range 96 - 97%, CoNSeP aug: median 97%, range 97 - 97%, MoNuSAC original: median 97%, range 97 - 98%, MoNuSAC aug: median 97%, range 97 - 98%), while precision was variable (PanNuke original: median 70%, range 10 - 80%, PanNuke aug: median 80%, range 10 - 85%, CoNSeP original: median 65%, range 20 - 85%, CoNSeP aug: median 70%, range 10 - 80%, MoNuSAC original: median 60%, range 30 - 60%, MoNuSAC aug: median 60%, range 30 - 60%). For inflammatory cells, recall was variable PanNuke original: median 40%, range 8 - 80%, PanNuke aug: median 45%, range 10 - 80%, CoNSeP original: median 43%, range 23 - 75%, CoNSeP aug: median 45%, range 10 - 90%, MoNuSAC original: median 50%, range 20 - 73%, MoNuSAC aug: median 58%, range 30 - 80%), while precision was > 90% for all models (PanNuke original: median 97%, range 97 - 97%, PanNuke aug: median 97%, range 97 - 97%, CoNSeP original: median 97%, range 89 - 97%, CoNSeP aug: median 97%, range 97 - 97%, MoNuSAC original: median 97%, range 97 - 97%, MoNuSAC aug: median 97%, range 97 - 97%). Figure 5 exemplifies the best, worst and largest range for classification of individual patches according to our visual validation. Both pathologists reported that for cancer cells, precision seemed to be directly correlated to the number of cancer cells, while for immune cells, recall seemed to be inversely correlated to the number of immune cells.

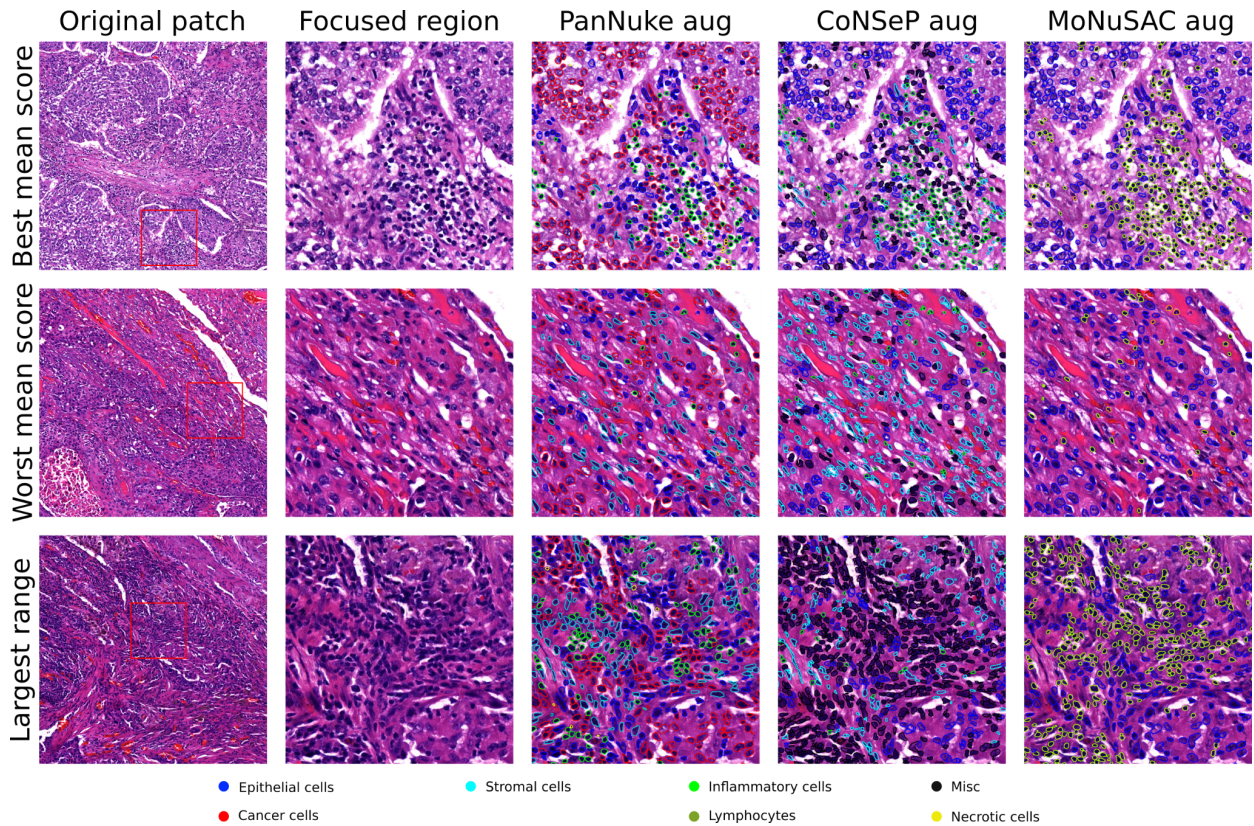

**Figure 5:** The top, middle and bottom rows represent the best, worst and largest range of F1 scores for immune cell classifications from our manual validation, respectively. The first column is an overview of the entire image patch, while the second to fifth column represents a focused region with and without detection/classification overlays.

# Discussion

## Summary of the results

In this paper we have presented our adaptation of HoVer-Net. Specifically, we have implemented the original approach as presented by Graham et al. and retrained the model on an expanded dataset published by Gamper et al (Graham *et al.*, 2019; Gamper *et al.*, 2020). To increase model accuracy and prevent overfitting, we introduced additional augmentations and compared models trained with and without these. Interestingly, our additional augmentations did not lead to a measurable increase in model performance (Tables 2 and S3). Moreover, we published UiT-TILs, a novel retrospective dataset consisting of patient information and tumor images sampled from 87 patients with NSCLC. We used this dataset to explore the possible clinical impact of our approach and compared it with current state-of-the-art methods for TIL evaluation in NSCLC patients. Our results suggest that the prognostic impact of TIL identification in H&E stained sections with the current DL approach is comparable, and possibly superior, to chromogenic assays (standard IHC stained CD8 cells in TMAs HR 0.34 95% CI 0.17-0.68 *vs* TILs in HE WSIs: HoVer-Net PanNuke Aug Model HR 0.30 95% CI 0.15-0.60 and HoVer-Net MoNuSAC Aug model HR 0.27 95% CI 0.14-0.53, Figure 4 and S5 Table). Finally, to provide an environment for easy and fast prototyping and deployment for experiments and production, we built a cloud based WSI viewer to support our backend. Our system is scalable and both the frontend and backend can function separately.

## Related works and lessons learned

The number of potential use-cases for digital pathology is increasing. However, while basic research is making strong headway, the use of digital pathology in routine clinical pathology is currently not aligned with expectations for most laboratories. In a recent paper, Hanna et al. reviews the process of adopting digital pathology in a clinical setting and shares from their process at the Department of Pathology, Memorial Sloan Kettering Cancer Center (Hanna *et al.*, 2021). They conclude that to increase the number of pathologists willing to adopt digital pathology, the technology has to offer benefits that impact the pathologists' daily work. While conventional image analyses have proven suitable for some tasks like counting the number of cancer cells staining positive for Ki67 in breast cancer (Goncalves *et al.*, 2017), they are prone to failure given increasing complexity. We, among many others, believe that DL will deliver the tools necessary to irreversibly pave the way for digital pathology. Despite this optimism, one of the main arguments against DL in medicine in general, and in cancer care in particular, is its inherent black box nature. As an example, Skrede et al. published DoMore-v1-CRC, a prognostic model in colorectal cancer that outperformed the prognostic impact of the TNM staging system for CRC patients in retrospective data (Skrede *et al.*, 2020). Moreover, their model may help select patients that will benefit from adjuvant chemotherapy. However, their lack of explainability may impede implementation. We chose to focus on cell segmentation and classification in general and on TILs in particular. Distinct from models that derive patients' prognoses from abstract features in imaging data alone, the success of our model can be evaluated directly based on its output (Figure 5). Interestingly, the prognostic impact derived from TILs identified on NSCLC H&E WSIs by current state of the art segmentation and classification models are equal or superior to IHC based methods. This latter finding was corroborated by our manual validation where segmentation and classification of TILs were consistent through the majority of images. However, we observed a significant difference in model performance for an image with a darker background and overall worse performance in images where TILs were scarce. These latter points suggest that our models are overfitting on the dataset. To summarize, our present effort illustrates that it is possible to repurpose existing DL algorithms for use on real patient samples with understandable output translating into clinical information that may be used to make informed treatment related decisions for patients.

In an attempt to further generalize our models, we adapted a linear augmentation policy proposed by Balkehol et al. to the HoVer-Net training pipeline (Balkenhol *et al.*, 2018). For models trained on the PanNuke and MoNuSAC datasets, we did not observe the expected benefit of the additional augmentations (Tables 2 and S3). For models trained on the CoNSeP dataset, there was a tendency towards better performance when additional augmentations were used. Currently, there is no reference augmentation policy for use on histopathology datasets. Our experience suggests that smaller datasets may benefit the most from additional augmentation. However, there are several steps including, but not limited to, tissue handling, staining, scanning, normalization and additional augmentations fitted before or during training that may impact model performance. Exploring these is beyond the scope of the current paper, but our use of MLOps best practices should make it easy to evaluate different normalization techniques and augmentation policies. Based on the reasonable performance of the CoNSeP models on CRC tissues and on the CoNSeP dataset, another feasible approach is to generate specific datasets for each problem. This latter strategy may make models less complex because variability will be reduced. However, we do not believe that task specific datasets will contribute to more accurate TIL detection as TILs, contrary to cancer cells, are morphologically equal across all cancers.

## Future work

As a direction for the future work we can outline the following points: prune and optimizing models to run faster with less computational power, improve web-server application to reduce response times, integrate with widely used tools such as ImageJ/Fiji, Cell Profiler, QuPath and others, add additional features useful as decision support for pathologists, and test acquired models on new datasets to validate model performance on other tissue types. It is also pertinent to explore new methods for dataset generation that incorporate all steps from tissue handling to staining to segmentation, auto-labeling and manual and/or automatic feedback.

## Limitations

In our composed dataset, we manually extracted patches to test our models and hypotheses. In future works, we plan to eliminate that flaw and automatically extract the region of interest in a WSI.

In this work, our aim was to provide an interactive graphical user interface for pathologists able to test and evaluate different machine learning models for WSI analysis. The response time is minutes for average size patches and hours for WSIs. It can be improved by using a more powerful GPU or multiple GPUs. However, we also want to make the service accessible to pathologists by using cloud services. The cloud service cost may be significant, especially if the service is running continuously. Model optimization may reduce cost and improve response times, for example using hardware accelerators, or using pruning techniques. It may also be more cost efficient to deploy the systems on an on-premise server with GPUs. Currently our tool is designed for hypothesis generation, testing and education. In a clinical implementation we need to improve response times, reduce costs, make it easier and more convenient to use, integrate with existing clinical ecosystems, test in clinical scenarios and obtain permissions from regulatory bodies for use as a medical device.

## Conclusion

In conclusion, we have used already published state-of-the-art DL models for segmentation and classification of cells, and adapted and retrained these for use on new data. We have shown that TIL identification using DL models on H&E WSIs is possible and further that the clinical results obtained are comparable to, and possibly supersedes, the prognostic impact of conventional IHC based methods for TIL identification in tissue from lung cancer patients. Moreover, our approach, utilizing published datasets, significantly limits the need of creating additional costly labeled datasets. Distributed deployment, and deployment as a service, allows for easier and faster model deployment, potentially facilitating the use of DL models in digital pathology. However, before clinical implementation, the efficacy of our methods needs to be further validated in independent datasets and the system must be tightly integrated into the clinical work-flow to ensure adaptation to daily diagnostics.

# Acknowledgments

This work was supported by the High North Population Studies, UiT The Arctic University of Norway. This work was funded in part by the Research Council of Norway grant no. 309439 SFI Visual Intelligence, and the North Norwegian Health Authority grant no. HNF1521-20.

# Supplementary tables

**S1 Table**: Clinicopathologic variables and their correlations with median dichotomized TILs identified using the PanNuke aug model (n = 87, chi-square and Fisher's exact tests)

|  | N | (%) | PanNuke score Low | High | P |
|---|---|---|---|---|---|
| Age |  |  |  |  | 0.72 |
| [0,65] | 33 | 38% | 18 | 15 |  |
| (65,85] | 54 | 62% | 26 | 28 |  |
| Gender |  |  |  |  | 0.57 |
| Female | 21 | 24% | 9 | 12 |  |
| Male | 66 | 76% | 35 | 31 |  |
| Smoking status |  |  |  |  | 0.76 |
| Never smoked | 7 | 8% | 4 | 3 |  |
| Present smoker | 45 | 52% | 24 | 21 |  |
| Previous smoker | 35 | 40% | 16 | 19 |  |
| ECOG |  |  |  |  | 0.48 |
| 0 | 45 | 52% | 20 | 25 |  |
| 1 | 34 | 39% | 20 | 14 |  |
| 2 | 8 | 9% | 4 | 4 |  |
| Histology |  |  |  |  | 0.10 |
| LUAD | 35 | 40% | 22 | 13 |  |
| LUSC | 52 | 60% | 22 | 30 |  |
| tStage |  |  |  |  | 0.43 |
| 1 | 24 | 28% | 15 | 9 |  |
| 2 | 37 | 43% | 18 | 19 |  |
| 3 | | 19 | 22% | 9 | 10 |
| 4 | | 7 | 8% | 2 | 5 |
| nStage | | | | | 0.28 |
| N0 | | 55 | 63% | 30 | 25 |
| N1 | | 22 | 25% | 8 | 14 |
| N2 | | 10 | 11% | 6 | 4 |
| pStage | | | | | 0.06 |
| I | | 33 | 38% | 22 | 11 |
| II | | 35 | 40% | 14 | 21 |
| III | | 19 | 22% | 8 | 11 |
| Differentiation | | | | | 0.47 |
| Well | | 10 | 11% | 7 | 3 |
| Moderate | | 34 | 39% | 16 | 18 |
| Poor | | 43 | 49% | 21 | 22 |

Abbreviations: ECOG, Eastern Collaborative Oncology Group; LUAD, lung adenocarcinoma; LUSC, lung squamous cell carcinoma.

S2 Table: HoVer-Net trained on the CoNSeP dataset (A I) and our reproduction (A II), and HoVer-Net trained on the PanNuke dataset (B I) and our replication (B II).

|  | A) | | B) | |
|---|---|---|---|---|
|  | I | II | I | II |
|  | CoNSeP original | CoNSeP reproduction | PanNuke | PanNuke reproduction |
| Reference | (Graham *et al.*, 2019) |  | (Gamper *et al.*, 2020) |  |
| *Segmentation* | | | | |
| Dice2 | 0.85 | 0.84 | NG | 0.83 |
| AJI | 0.57 | 0.53 | NG | 0.67 |
| DQ | 0.70 | 0.66 | NG | 0.78 |
| SQ | 0.78 | 0.77 | NG | 0.61 |
| PQ | 0.55 | 0.50 | NG | 0.78 |
| AJI+ | NG | 0.55 | NG | 0.68 |
| *Integrated Classification* | | | | |
| $F1_{overall}$ | 0.75 | 0.75 | NG | 0.80 |
| $Accuracy_{overall}$ | NG | 0.83 | NG | 0.76 |
| $Precision_{overall}$ | NG | 0.74 | NG | 0.79 |
| $Recall_{overall}$ | NG | 0.75 | NG | 0.80 |
| $F1_{inflammatory}$ | 0.63 | 0.52 | 0.54 | 0.50 |
| $Precision_{inflammatory}$ | NG | 0.73 | 0.56 | 0.48 |
| $Recall_{inflammatory}$ | NG | 0.40 | 0.51 | 0.53 |
| $F1_{cancer}$ | 0.64 | 0.66 | 0.62 | 0.58 |
| $Precision_{cancer}$ | NG | 0.59 | 0.58 | 0.57 |
| $Recall_{cancer}$ | NG | 0.74 | 0.67 | 0.60 |

S3 Table: A summary of the performance of four deep learning models trained using the CoNSeP (A I-II and B I-II) and the PanNuke (A III-IV and B III-IV) datasets using the original training pipeline as published by Graham et al. (Graham *et al.*, 2019) without (A I and III and B I and III) and with (A II and IV and B II and IV) enhanced augmentation (Graham *et al.*, 2019). The best results for each parameter 1) within each dataset are in bold and 2) for models trained on another dataset are in italics. Integrated classification results are in Table 2.

| Test data | CoNSeP | | | | | | PanNuke | | | | | | MoNuSAC | | | | | |
|---|---|---|---|---|---|---|---|---|---|---|---|---|---|---|---|---|---|---|
| Model | CoNSeP | | PanNuke | | MoNuAC | | CoNSeP | | PanNuke | | MoNuSAC | | CoNSeP | | PanNuke | | MoNuAC | |
| Augmentation | HoVer | Aug | HoVer | Aug | HoVer | Aug | HoVer | Aug | HoVer | Aug | HoVer | Aug | HoVer | Aug | HoVer | Aug | HoVer | Aug |
| Numbering | AI | AII | AIII | AIV | AV | AVI | BI | BII | BIII | BIV | BV | BVI | CI | CII | CIII | CIV | CV | CVI |
| Segmentation | | | | | | | | | | | | | | | | | | |
| Dice2 | **0.84** | **0.84** | *0.83* | 0.82 | 0.56 | 0.55 | 0.62 | *0.70* | **0.83** | **0.83** | 0.60 | 0.62 | 0.50 | 0.61 | 0.70 | *0.71* | 0.74 | **0.75** |
| AJI | **0.53** | **0.53** | *0.52* | 0.51 | 0.32 | 0.32 | 0.42 | *0.47* | 0.67 | **0.68** | 0.45 | 0.46 | 0.33 | 0.41 | *0.52* | *0.52* | 0.56 | **0.57** |
| DQ | **0.66** | 0.65 | 0.63 | *0.64* | 0.41 | 0.41 | 0.50 | 0.55 | **0.78** | **0.78** | 0.57 | *0.58* | 0.42 | 0.50 | 0.68 | *0.69* | 0.75 | **0.76** |
| SQ | **0.77** | 0.50 | 0.48 | *0.49* | 0.32 | 0.32 | 0.32 | 0.38 | 0.61 | **0.62** | 0.41 | 0.42 | 0.30 | 0.38 | 0.52 | *0.53* | 0.60 | **0.61** |
| PQ | 0.50 | 0.77 | 0.76 | 0.77 | **0.78** | **0.78** | 0.63 | 0.68 | 0.78 | **0.79** | *0.70* | *0.70* | 0.61 | 0.70 | 0.75 | *0.76* | 0.79 | **0.80** |
| AJI+ | 0.55 | 0.55 | 0.55 | *0.56* | 0.35 | 0.35 | 0.42 | *0.48* | 0.68 | **0.69** | 0.45 | 0.47 | 0.33 | 0.41 | 0.52 | *0.53* | 0.57 | **0.58** |
| $Recall_d^{inflammatory}$ | 0.82 | 0.84 | ***0.89*** | 0.88 | 0.70 | 0.71 | *0.91* | 0.73 | 0.85 | 0.84 | 0.74 | 0.78 | 0.84 | 0.90 | 0.91 | *0.93* | 0.92 | 0.92 |
| $Recall_d^{cancer}$ | **0.78** | **0.78** | *0.74* | 0.72 | 0.54 | 0.52 | 0.45 | 0.58 | 0.82 | **0.83** | *0.62* | *0.62* | 0.24 | 0.81 | ***0.97*** | ***0.97*** | 0.87 | 0.89 |
| Classification | | | | | | | | | | | | | | | | | | |
| $Accuracy_c^{inflammatory}$ | 0.61 | 0.67 | 0.61 | 0.65 | 0.71 | **0.72** | 0.29 | 0.35 | 0.57 | **0.59** | *0.48* | 0.47 | 0.27 | 0.49 | *0.79* | 0.75 | **0.87** | **0.87** |
| $Precision_c^{inflammatory}$ | **0.85** | 0.83 | 0.64 | 0.73 | *0.84* | 0.81 | *0.66* | 0.59 | 0.70 | **0.76** | 0.56 | 0.51 | *0.94* | 0.91 | 0.92 | 0.92 | 0.92 | 0.91 |
| $Recall_c^{inflammatory}$ | 0.69 | 0.78 | ***0.93*** | 0.86 | 0.83 | 0.87 | 0.35 | 0.46 | 0.76 | 0.72 | 0.78 | **0.85** | 0.28 | 0.52 | *0.84* | 0.80 | 0.94 | **0.95** |
| $F1_c^{inflammatory}$ | 0.76 | 0.80 | 0.76 | 0.79 | 0.83 | **0.84** | 0.45 | 0.51 | 0.73 | **0.74** | *0.65* | 0.64 | 0.43 | 0.66 | 0.88 | *0.86* | **0.93** | **0.93** |
| $Accuracy_c^{cancer}$ | 0.88 | **0.89** | *0.81* | 0.79 | 0.70 | 0.73 | 0.38 | 0.44 | 0.72 | **0.73** | *0.69* | 0.68 | 0.15 | 0.45 | *0.75* | 0.73 | **0.85** | **0.85** |
| $Precision_c^{cancer}$ | 0.90 | **0.93** | *0.86* | 0.81 | 0.72 | 0.76 | 0.89 | *0.92* | 0.83 | 0.80 | 0.80 | 0.83 | *0.94* | *0.94* | 0.78 | 0.75 | 0.87 | 0.88 |
| $Recall_c^{cancer}$ | **0.97** | 0.94 | 0.93 | *0.97* | 0.96 | 0.95 | 0.40 | 0.46 | 0.83 | **0.89** | *0.83* | 0.79 | 0.15 | 0.46 | *0.96* | 0.96 | **0.98** | 0.97 |
| $F1_c^{cancer}$ | **0.94** | **0.94** | *0.89* | 0.88 | 0.83 | 0.84 | 0.55 | 0.61 | 0.83 | **0.84** | *0.82* | 0.80 | 0.27 | 0.62 | *0.86* | 0.85 | **0.92** | **0.92** |

S4 Table: A comparison of disease-specific survival of NSCLC patients according to high and low levels of TILs identified in H&E WSIs using different approaches (n = 87, univariable analyses, log-rank test).

| | N(%) | 5 Year | Median | HR(95%CI) | P |
|---|---|---|---|---|---|
| TILs in H&E WSIs: manual score | | | | | 0.173 |
|   0-5% | 9(10) | 44 | 27 | 1.000 | |
|   6-25% | 47(54) | 51 | 114 | 0.8(0.24-2.66) | |
|   26-50% | 22(25) | 76 | NA | 0.39(0.11-1.33) | |
|   >50% | 5(6) | 80 | NA | 0.25(0.05-1.28) | |
|   Missing | 4(5) | | | | |
| CD8+ cells in DAB stained TMAs: QuPath cell count | | | | | **0.003** |
|   Low | 41(47) | 44 | 51 | 1.000 | |
|   High | 41(47) | 79 | NA | 0.34(0.17-0.68) | |
|   Missing | 5(6) | | | | |
| TILs in H&E WSIs: rule based approach | | | | | 0.095 |
|   Low | 44(51) | 54 | 114 | 1.000 | |
|   High | 43(49) | 69 | NA | 0.56(0.29-1.1) | |
| TILs in HE WSIs: HoVer-Net CoNSeP Aug Model | | | | | **0.013** |
|   Low | 44(51) | 51 | 64 | 1.000 | |
|   High | 43(49) | 70 | NA | 0.42(0.22-0.84) | |
| TILs in HE WSIs: HoVer-Net CoNSeP Orig Model | | | | | 0.088 |
|   Low | 44(51) | 55 | 83 | 1.000 | |
|   High | 43(49) | 67 | NA | 0.55(0.28-1.08) | |
| TILs in HE WSIs: HoVer-Net MoNuSAC Aug Model | | | | | **<0.001** |
|   Low | 44(51) | 44 | 47 | 1.000 | |
|   High | 43(49) | 77 | NA | 0.27(0.14-0.53) | |
| TILs in HE WSIs: HoVer-Net MoNuSAC Orig Model | | | | | **<0.001** |
|   Low | 44(51) | 48 | 47 | 1.000 | |
|   High | 43(49) | 73 | NA | 0.35(0.18-0.69) | |
| TILs in HE WSIs: HoVer-Net PanNuke Aug Model | | | | | **<0.001** |
|   Low | 44(51) | 44 | 47 | 1.000 | |
|   High | 43(49) | 77 | NA | 0.3(0.15-0.6) | |
| TILs in HE WSIs: HoVer-Net PanNuke Orig Model | | | | | **0.007** |
|   Low | 44(51) | 49 | 51 | 1.000 | |
|   High | 43(49) | 72 | NA | 0.39(0.2-0.76) | |

Abbreviations: TIL, tissue infiltrating lymphocyte; CD, cluster of differentiation;

S5 Table: Manual estimation of precision and recall and calculated F1 scores for cancer- and immune cells on the output of the PanNuKe, CoNSeP and MoNuSAC models with updated augmentation. The 20 1000x1000µm patches were randomly sampled from our lung cancer cohort.

| | PanNuke aug | | | | | | CoNSeP aug | | | | | | MoNuSAC aug | | | | | |
|---|---|---|---|---|---|---|---|---|---|---|---|---|---|---|---|---|---|---|
| | Cancer cells | | | Immune cells | | | Cancer cells | | | Immune cells | | | Cancer cells | | | Immune cells | | |
| Slide | Rec | Prec | F1 | Rec | Pre | F1 | Rec | Prec | F1 | Rec | Pre | F1 | Rec | Prec | F1 | Rec | Pre | F1 |
| 1 | 0.97 | 0.80 | 0.88 | 0.80 | 0.97 | 0.88 | 0.97 | 0.80 | 0.88 | 0.90 | 0.97 | 0.93 | 0.97 | 0.90 | 0.93 | 0.80 | 0.97 | 0.88 |
| 2 | 0.97 | 0.80 | 0.88 | 0.80 | 0.97 | 0.88 | 0.97 | 0.80 | 0.88 | 0.90 | 0.97 | 0.93 | 0.98 | 0.75 | 0.85 | 0.80 | 0.97 | 0.88 |
| 3 | 0.97 | 0.85 | 0.91 | 0.70 | 0.97 | 0.81 | 0.97 | 0.80 | 0.88 | 0.80 | 0.97 | 0.88 | 0.97 | 0.90 | 0.93 | 0.80 | 0.97 | 0.88 |
| 4 | 0.97 | 0.80 | 0.88 | 0.70 | 0.97 | 0.81 | 0.97 | 0.80 | 0.88 | 0.80 | 0.97 | 0.88 | 0.97 | 0.80 | 0.88 | 0.80 | 0.97 | 0.88 |
| 5 | 0.97 | 0.30 | 0.46 | 0.50 | 0.97 | 0.66 | 0.97 | 0.30 | 0.46 | 0.50 | 0.97 | 0.66 | 0.97 | 0.55 | 0.67 | 0.80 | 0.97 | 0.88 |
| 6 | 0.97 | 0.30 | 0.46 | 0.10 | 0.97 | 0.18 | 0.97 | 0.20 | 0.33 | 0.10 | 0.97 | 0.18 | 0.97 | 0.60 | 0.74 | 0.70 | 0.97 | 0.81 |
| 7 | 0.97 | 0.10 | 0.18 | 0.20 | 0.97 | 0.33 | 0.97 | 0.10 | 0.18 | 0.30 | 0.97 | 0.46 | 0.97 | 0.30 | 0.46 | 0.60 | 0.97 | 0.74 |
| 8 | | | | 0.20 | 0.97 | 0.17 | | | | 0.30 | 0.97 | 0.46 | | | | 0.50 | 0.97 | 0.66 |
| 9 | 0.97 | 0.30 | 0.46 | 0.30 | 0.97 | 0.46 | 0.97 | 0.30 | 0.46 | 0.30 | 0.97 | 0.46 | 0.97 | 0.50 | 0.66 | 0.50 | 0.97 | 0.66 |
| 10 | 0.97 | 0.80 | 0.88 | 0.80 | 0.97 | 0.88 | 0.97 | 0.80 | 0.88 | 0.70 | 0.97 | 0.81 | 0.97 | 0.90 | 0.93 | 0.55 | 0.97 | 0.70 |
| 11 | 0.97 | 0.50 | 0.66 | 0.50 | 0.97 | 0.66 | 0.97 | 0.50 | 0.66 | 0.50 | 0.97 | 0.66 | 0.97 | 0.50 | 0.66 | 0.40 | 0.97 | 0.57 |
| 12 | 0.97 | 0.60 | 0.74 | 0.50 | 0.97 | 0.66 | 0.97 | 0.50 | 0.66 | 0.40 | 0.97 | 0.57 | 0.97 | 0.40 | 0.57 | 0.40 | 0.97 | 0.57 |
| 13 | 0.97 | 0.70 | 0.81 | 0.40 | 0.97 | 0.57 | 0.97 | 0.70 | 0.81 | 0.40 | 0.97 | 0.57 | 0.97 | 0.75 | 0.85 | 0.80 | 0.97 | 0.88 |
| 14 | 0.97 | 0.80 | 0.88 | 0.60 | 0.97 | 0.74 | 0.97 | 0.80 | 0.88 | 0.60 | 0.97 | 0.74 | 0.97 | 0.75 | 0.85 | 0.80 | 0.97 | 0.88 |
| 15 | 0.97 | 0.80 | 0.88 | 0.40 | 0.97 | 0.57 | 0.97 | 0.70 | 0.81 | 0.30 | 0.97 | 0.46 | 0.97 | 0.55 | 0.69 | 0.50 | 0.97 | 0.66 |
| 16 | 0.97 | 0.80 | 0.88 | 0.40 | 0.97 | 0.57 | 0.97 | 0.70 | 0.81 | 0.50 | 0.97 | 0.66 | 0.97 | 0.50 | 0.66 | 0.40 | 0.97 | 0.57 |
| 17 | 0.97 | 0.80 | 0.88 | 0.80 | 0.97 | 0.88 | 0.97 | 0.80 | 0.88 | 0.70 | 0.97 | 0.81 | 0.97 | 0.90 | 0.93 | 0.80 | 0.97 | 0.88 |
| 18 | 0.97 | 0.70 | 0.81 | 0.20 | 0.97 | 0.33 | 0.97 | 0.70 | 0.81 | 0.20 | 0.97 | 0.33 | 0.97 | 0.40 | 0.57 | 0.30 | 0.97 | 0.46 |
| 19 | 0.97 | 0.70 | 0.81 | 0.30 | 0.97 | 0.46 | 0.97 | 0.70 | 0.81 | 0.30 | 0.97 | 0.46 | 0.97 | 0.40 | 0.57 | 0.40 | 0.97 | 0.57 |
| 20 | 0.97 | 0.80 | 0.88 | 0.30 | 0.97 | 0.46 | 0.97 | 0.70 | 0.81 | 0.30 | 0.97 | 0.46 | 0.97 | 0.60 | 0.74 | 0.30 | 0.97 | 0.46 |
| Med | 0.97 | *0.64* | **0.75** | *0.48* | 0.97 | *0.60* | 0.97 | 0.62 | 0.72 | *0.49* | 0.97 | *0.62* | 0.97 | *0.63* | 0.74 | **0.60** | 0.97 | **0.72** |
| Mean | 0.97 | *0.80* | **0.88** | *0.45* | 0.97 | *0.61* | 0.97 | *0.70* | *0.81* | *0.45* | 0.97 | *0.61* | 0.97 | 0.60 | 0.74 | **0.58** | 0.97 | **0.72** |
| Min | 0.97 | 0.10 | 0.18 | *0.10* | 0.97 | *0.17* | 0.97 | 0.10 | 0.18 | 0.10 | 0.97 | 0.18 | 0.97 | 0.30 | **0.46** | **0.30** | 0.97 | **0.46** |
| Max | 0.97 | *0.85* | *0.91* | 0.80 | 0.97 | 0.88 | 0.97 | 0.80 | 0.88 | **0.90** | 0.97 | **0.93** | 0.98 | 0.90 | **0.93** | *0.80* | 0.97 | *0.88* |
| Range | 0.00 | *0.75* | *0.72* | *0.70* | 0.00 | *0.71* | 0.00 | 0.70 | 0.70 | 0.80 | 0.00 | 0.75 | 0.01 | **0.60** | **0.48** | **0.50** | 0.00 | **0.42** |

S6 Table: Manual estimation of precision and recall and calculated F1 scores for cancer- and immune cells on the output of the PanNuKe, CoNSeP and MoNuSAC models with original augmentation. The 20 1000x1000µm patches were randomly sampled from our lung cancer cohort.

| | PanNuke original | | | | | | CoNSeP original | | | | | | MoNuSAC original | | | | | |
|---|---|---|---|---|---|---|---|---|---|---|---|---|---|---|---|---|---|---|
| | Cancer cells | | | Immune cells | | | Cancer cells | | | Immune cells | | | Cancer cells | | | Immune cells | | |
| Slide | Rec | Prec | F1 | Rec | Pre | F1 | Rec | Prec | F1 | Rec | Pre | F1 | Rec | Prec | F1 | Rec | Pre | F1 |
| 1 | 0.96 | 0.80 | 0.87 | 0.76 | 0.97 | 0.85 | 0.97 | 0.85 | 0.91 | 0.70 | 0.97 | 0.81 | 0.97 | 0.85 | 0.91 | 0.70 | 0.97 | 0.81 |
| 2 | 0.96 | 0.80 | 0.87 | 0.80 | 0.97 | 0.88 | 0.97 | 0.79 | 0.87 | 0.60 | 0.97 | 0.72 | 0.98 | 0.75 | 0.85 | 0.60 | 0.97 | 0.74 |
| 3 | 0.96 | 0.78 | 0.85 | 0.70 | 0.97 | 0.81 | 0.97 | 0.85 | 0.91 | 0.70 | 0.89 | 0.78 | 0.97 | 0.90 | 0.93 | 0.70 | 0.97 | 0.81 |
| 4 | 0.96 | 0.80 | 0.87 | 0.75 | 0.97 | 0.85 | 0.97 | 0.80 | 0.88 | 0.75 | 0.97 | 0.85 | 0.97 | 0.70 | 0.81 | 0.65 | 0.97 | 0.78 |
| 5 | 0.97 | 0.29 | 0.44 | 0.40 | 0.97 | 0.57 | 0.97 | 0.60 | 0.72 | 0.45 | 0.97 | 0.53 | 0.97 | 0.55 | 0.67 | 0.70 | 0.97 | 0.81 |
| 6 | 0.97 | 0.20 | 0.33 | 0.10 | 0.97 | 0.18 | 0.96 | 0.40 | 0.54 | 0.40 | 0.97 | 0.50 | 0.97 | 0.60 | 0.74 | 0.65 | 0.97 | 0.78 |
| 7 | 0.97 | 0.10 | 0.18 | 0.08 | 0.97 | 0.14 | 0.97 | 0.20 | 0.32 | 0.40 | 0.97 | 0.54 | 0.97 | 0.30 | 0.46 | 0.50 | 0.97 | 0.66 |
| 8 | | | | 0.20 | 0.97 | 0.17 | | | | 0.50 | 0.97 | 0.66 | | | | 0.40 | 0.97 | 0.57 |
| 9 | 0.97 | 0.20 | 0.33 | 0.30 | 0.97 | 0.46 | 0.97 | 0.40 | 0.56 | 0.35 | 0.97 | 0.50 | 0.97 | 0.50 | 0.66 | 0.45 | 0.97 | 0.61 |
| 10 | 0.97 | 0.80 | 0.88 | 0.80 | 0.97 | 0.88 | 0.97 | 0.85 | 0.91 | 0.60 | 0.97 | 0.74 | 0.97 | 0.80 | 0.88 | 0.50 | 0.97 | 0.65 |
| 11 | 0.97 | 0.45 | 0.61 | 0.50 | 0.97 | 0.66 | 0.97 | 0.45 | 0.61 | 0.35 | 0.97 | 0.51 | 0.97 | 0.50 | 0.66 | 0.30 | 0.97 | 0.46 |
| 12 | 0.97 | 0.50 | 0.65 | 0.50 | 0.97 | 0.66 | 0.97 | 0.40 | 0.57 | 0.25 | 0.97 | 0.37 | 0.97 | 0.40 | 0.57 | 0.30 | 0.97 | 0.46 |
| 13 | 0.96 | 0.70 | 0.81 | 0.40 | 0.97 | 0.57 | 0.97 | 0.70 | 0.81 | 0.58 | 0.97 | 0.70 | 0.97 | 0.70 | 0.81 | 0.65 | 0.97 | 0.78 |
| 14 | 0.96 | 0.80 | 0.87 | 0.50 | 0.97 | 0.66 | 0.97 | 0.75 | 0.85 | 0.45 | 0.97 | 0.53 | 0.97 | 0.75 | 0.85 | 0.73 | 0.97 | 0.83 |
| 15 | 0.97 | 0.80 | 0.88 | 0.40 | 0.97 | 0.57 | 0.97 | 0.55 | 0.69 | 0.38 | 0.97 | 0.53 | 0.97 | 0.55 | 0.69 | 0.40 | 0.97 | 0.57 |
| 16 | 0.97 | 0.80 | 0.88 | 0.35 | 0.97 | 0.51 | 0.97 | 0.65 | 0.77 | 0.30 | 0.97 | 0.45 | 0.97 | 0.50 | 0.66 | 0.35 | 0.97 | 0.51 |
| 17 | 0.97 | 0.80 | 0.88 | 0.80 | 0.97 | 0.88 | 0.97 | 0.85 | 0.91 | 0.75 | 0.97 | 0.85 | 0.97 | 0.80 | 0.88 | 0.70 | 0.97 | 0.81 |
| 18 | 0.97 | 0.70 | 0.81 | 0.20 | 0.97 | 0.33 | 0.97 | 0.55 | 0.69 | 0.23 | 0.97 | 0.36 | 0.97 | 0.40 | 0.57 | 0.20 | 0.97 | 0.33 |
| 19 | 0.97 | 0.70 | 0.81 | 0.30 | 0.97 | 0.46 | 0.97 | 0.55 | 0.69 | 0.25 | 0.97 | 0.37 | 0.97 | 0.40 | 0.57 | 0.35 | 0.97 | 0.51 |
| 20 | 0.97 | 0.70 | 0.81 | 0.33 | 0.97 | 0.49 | 0.97 | 0.70 | 0.81 | 0.25 | 0.97 | 0.39 | 0.97 | 0.60 | 0.74 | 0.20 | 0.97 | 0.33 |
| Med | 0.97 | 0.62 | 0.72 | 0.46 | 0.97 | 0.58 | 0.97 | 0.63 | 0.74 | 0.46 | 0.97 | 0.58 | 0.97 | 0.61 | 0.73 | 0.50 | 0.97 | 0.64 |
| Mean | 0.97 | 0.70 | 0.81 | 0.40 | 0.97 | 0.57 | 0.97 | 0.65 | 0.77 | 0.43 | 0.97 | 0.53 | 0.97 | 0.60 | 0.74 | 0.50 | 0.97 | 0.66 |
| Min | 0.96 | 0.10 | 0.18 | 0.08 | 0.97 | 0.14 | 0.96 | 0.20 | 0.32 | 0.23 | 0.89 | 0.36 | 0.97 | 0.30 | 0.46 | 0.20 | 0.97 | 0.33 |
| Max | 0.97 | 0.80 | 0.88 | 0.80 | 0.97 | 0.88 | 0.97 | 0.85 | 0.91 | 0.75 | 0.97 | 0.85 | 0.98 | 0.90 | **0.93** | 0.73 | 0.97 | 0.83 |
| Range | 0.02 | 0.70 | 0.70 | 0.73 | 0.00 | 0.74 | 0.01 | 0.65 | 0.59 | 0.53 | 0.09 | 0.49 | 0.01 | **0.60** | **0.48** | 0.53 | 0.00 | 0.50 |

# Supplementary figures

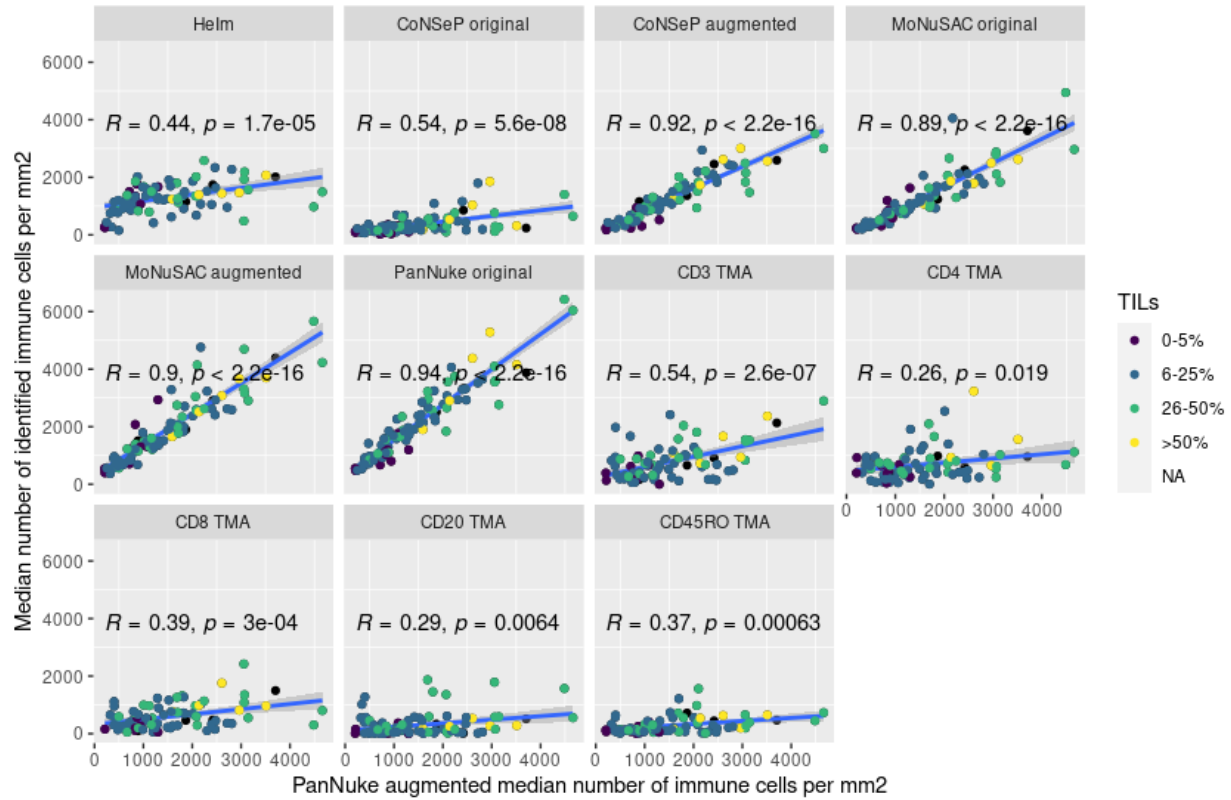

**S1 Figure:** Median number of immune cells for each patient identified using the augmented PanNuke model plotted against the median number of immune cells identified by: 1) Helm, CoNSeP original (HoVer-Net paper), CoNSeP augmented, MoNuSAC original, MoNuSAC augmented and PanNuke original, using the dataset provided in this publication and 2) CD3, CD4, CD8, CD20 and CD45RO identified on TMAs for each patient (the full TMA results are published (Kilvaer *et al.*, 2020). Each dot represents data from an individual patient and is labelled according to the semi-quantitative TILs score he/she was assigned in (Rakaee *et al.*, 2018).

# Supplementary materials and methods

## Helm: A simple rule based algorithm

We implement a simple rule based approach for annotating cells as a baseline for automated approaches. We first do simple image augmentation by converting from the RGB to the HED color space using color deconvolution as described above. Then we generate a mask thresholding each channel in the HED color space. The mask is then morphologically opened using an elliptical and then a square kernel. This results in two masks.

We apply c*ontour detection* and filtering to ensure that only detections within the tolerated size and circularity range are kept. We remove duplicates using overlap detection. Finally, we set the ranges of the hematoxylin and eosin image channel for the mask to 220-255 and 0-50, the range of the area to 190-600px$^2$ and the minimum circularity to 65. The source code is in: https://github.com/uit-hdl/HEImmune.

## Definition of metrics used in the replication study

To compare our results with those previously published by Graham et al, (Graham *et al.*, 2019) we used the custom defined F-score for different types of cells they propose in their paper.

$$F^t_{dcr} = \frac{2(TP^t_{dcr} + TN^t_{dcr})}{2(TP^t_{dcr} + TN^t_{dcr}) + \alpha_0 FP^t_{dcr} + \alpha_1 FN^t_{dcr} + \alpha_2 FP_d + \alpha_3 FN_d}$$

where $TP^t_{dcr}$ is correct instances classifications of type *t*, $TN^t_{dcr}$ as correctly classified instances of types other than type *t*, $FN^t_{dcr}$ as incorrectly classified instances of types other than type *t*, $FP^t_{dcr}$ as incorrectly classified instances of type t, $FN_d$ as misdetected GT instances and $FP_d$ as overdetected predicted instances, $\alpha_0 = \alpha_1 = 2$ and $\alpha_2 = \alpha_3 = 1$ for taking into account detection results, with higher emphasis on classification performance.

In order to get a deeper understanding of the models performance we introduced precision ($Precision^t_{dcr}$) and recall ($Recall^t_{dcr}$) scores, based on the custom defined F-score defined above, that encompass detection ($X_d$) and classification ($X_c$) results. We use $TPN = TP^t_{dcr} + TN^t_{dcr}$, $FP = 2 \cdot FP^t_{dcr} + FP_d$, $FN = 2 \cdot FN^t_{dcr} + FN_d$, to calculate $Precision^t_{dcr} = \frac{TPN}{TPN + FP}$ and $Recall^t_{dcr} = \frac{TPN}{TPN + FN}$. This latter approach is similar to Gamper et al. (Gamper *et al.*, 2020) and allows comparisons with models they built using the HoVer-Net infrastructure and PanNuke data.

## Definition of metrics used to separately evaluate segmentation and classification

Segmentation metrics include:

$DICE = \frac{2TP}{2TP + FP + FN}$, $DICE2$ computes and aggregates $DICE$ per nucleus.

$AJI = \frac{\sum_{i=1}^{N} |G_i \cap P_i|}{\sum_{i=1}^{N} |G_i \cap P_i| + \sum_{i \in rest} |P_i|}$, where $P_i$ is the predicted nucleus that maximizes the Jaccard Index with the ground truth nucleus $G_i$ and $rest$ refers to the collection of $P_i$ with no match.

$DQ = \frac{|TP|}{|TP| + 0.5\,|FP| + 0.5\,|FN|}$

$SQ = \frac{\sum_{(x,y) \in TP} IOU(x,y)}{|TP|}$, $IOU(x, y)$ - intersection over union between x and y sets

$PQ = DQ \times SQ$

$Recall_d^t = \frac{TP_d^t}{TP_d^t + FN_d^t}$, where $TP_d^t$ - detected cells with GT label t, $FN_d^t$ - cell with GT label t not detected.

Classification metrics include:

$Accuracy_c^t = \frac{TP_c^t + TN_c^t}{TP_c^t + TN_c^t + FP_c^t + FN_c^t}$, $Precision_c^t = \frac{TP_c^t}{TP_c^t + FP_c^t}$, $Recall_c^t = \frac{TP_c^t}{TP_c^t + FN_c^t}$, $F1_c^t = \frac{TP_c^t}{TP_c^t + 0.5(FP_c^t + FN_c^t)}$,

where

$TP_c^t$ - detected cells with GT label t, classified as t,

$TN_c^t$ - detected cells with GT label other than t, classified as other than t,

$FP_c^t$ - detected cells with GT label other than t, classified as t,

$FN_c^t$ - detected cells with GT label t classified as other than t, for each class t.

# Supplementary results

## Model training and inference times

Training times for the CoNSeP, PanNuke, and MoNuSAC datasets were 30h, 60h and 31h on the Azure cloud and 11h, 38h and 10h on the local workstation, respectively. The cloud enables horizontal scaling for the inference part, thus time for the post processing step may be reduced, depending on available resources. Inference times on the UiT-TILs dataset were 27h and 12h corresponding to 81s and 36s for single patches, for Azure and workstation, respectively. Based on the size of the patches, we estimate that inference and post processing time for typical WSIs (10x15mm) is between 1h30min and 3h30min on our current Azure and workstation deployments.